\documentclass[runningheads]{llncs}
\usepackage{graphicx}
\usepackage{enumitem}
\usepackage{subcaption}
\usepackage{algorithm}
\usepackage{xcolor}
\usepackage{amssymb,amsmath,amsfonts}
\usepackage{wrapfig,booktabs}
\usepackage{lipsum,mwe}
\begin{document}
\title{Constraints Satisfiability Driven Reinforcement Learning for Autonomous Cyber Defense\thanks{Accepted at International Conference on Autonomous Intelligent Cyber-defence Agents (AICA 2021).}}
%
\titlerunning{Constraints Satisfiability Driven Reinforcement Learning (CSRL)}
%
\vspace{-2em}
\author{Ashutosh Dutta\inst{1}\and
Ehab Al-Shaer\inst{2}\and
Samrat Chatterjee \inst{3}}
\authorrunning{A. Dutta et al.}
%
\institute{
University of North Carolina Charlotte, Charlotte, NC, USA\\
\and
Carnegie Mellon University, Pittsburgh, PA, USA\\
\and
Pacific Northwest National Laboratory, Richland, WA 99354\\
\email{adutta6@uncc.edu}\inst{1},
\email{ehab@cmu.edu}\inst{2},
\email{samrat.chatterjee@pnnl.gov}\inst{3}}


%
\maketitle              
\vspace{-2em}
\begin{abstract}
With the increasing system complexity and attack sophistication, the necessity of autonomous cyber defense becomes vivid for cyber and cyber-physical systems (CPSs). Many existing frameworks in the current-state-of-the-art either rely on static models with unrealistic assumptions, or fail to satisfy the system safety and security requirements. In this paper, we present a new hybrid autonomous agent architecture that aims to optimize and verify defense policies of reinforcement learning (RL) by incorporating constraints verification (using satisfiability modulo theory (SMT)) into the agent's decision loop. The incorporation of SMT does not only ensure the satisfiability of safety and security requirements, but also provides constant feedback to steer the RL decision-making toward safe and effective actions. This approach is critically needed for CPSs that exhibit high risk due to safety or security violations. 
Our evaluation of the presented approach in a simulated CPS environment shows that the agent learns the optimal policy fast and defeats diversified attack strategies in 99\% cases.
\end{abstract}
\vspace{-1em}
\vspace{-2em}
\section{Introduction}
With wide applications spanning from national critical infrastructures (e.g., smart grid, transport) to personal domains (e.g., home automation system, healthcare), cyber and CPS systems become more susceptible to cyber attacks due to misconfigurations, unpatched, or unknown vulnerabilities. Moreover, attacks like Advanced Persistent Threat (APT) are well-resourced and highly sophisticated to cause serious and large damage for critical infrastructures within relatively a short time \cite{xiao2018dynamic}. 
Therefore, automating proactive defense such as penetration testing and risk identification, and reactive defense such as intrusion response is a key to maintain the integrity and security of these systems.

Developing autonomous agents for cyber defense is one of the most promising solutions to achieve real-time monitoring and response against advanced attackers with minimal human involvement. Autonomous cyber defense agents (ACDA) have capabilities to not only timely respond to malicious actions but also adapt their decision-making dynamically to cope with changes of environment or attack strategies. On the other hand, to guarantee the mission safety, ACDA actions must be shown provably correct according to the mission, operation, and business requirements. 

Researchers have applied game theory \cite{do2017game}, sequential decision process \cite{hu2017online,miehling2018pomdp}, and reinforcement learning \cite{panfili2018game,nguyen2019deep} to optimize defense response planning. However, these works have limited real-world applications due to struggling to converge while having numerous requirements. Several works apply constraint satisfaction problems (CSP) \cite{xiao2018dynamic,wang2013cyber} to optimize planning considering all requirements as constraints. However, these works rely on static models for critical parameters which may be very hard to formulate, even probabilistically, due to lack of domain specific data. Moreover, static assumptions on attackers' exploitation capabilities restrict attack behavior unrealistically. Therefore, current state-of-the-art of autonomous cyber defense lacks a framework that can optimize defense planning at real-time while satisfying all various requirements.

In this paper, we present a new hybrid autonomous agent architecture that optimizes defense policies through incorporating the feedback on constraint satisfiability into the decision loop.
We formulate the defense optimization problem as a Sequential Decision Process (SDP) \cite{sutton2018reinforcement}, where defense effectiveness depends on stochastic environment behavior, adaptive attack strategies, and mission-oriented safety and security requirements.
However, ACDA usually lacks domain-specific experience and data to predict attack behaviors or characterize defense effectiveness. 
To accomplish this goal, we develop a novel approach, named Constrained Satisfiability-driven Reinforcement Learning (CSRL) approach, to solve the SDP through learning the environment based on interactive experience with the environment. CSRL employs model-free Reinforcement Learning \cite{sutton2018reinforcement} to optimize the defense decision-making, and applies Satisfiability Modulo Theory (SMT)~\cite{barrett2018satisfiability} for constraints satisfiability verification to provide verifiability and refinement of the defense actions according to safety and security properties. The incorporation of SMT architecture guides the agent's RL algorithm towards safe and effective defense planning.

Our CSRL approach decouples the policy optimization and constraint satisfying modules to address the challenge of computation complexity. Instead of feeding constraints directly to the optimizer, the policy is updated based on the satisfiability of current constraint set by computed defense actions. This approach does not only make the agent computationally feasible for real-time defense optimization in a constrained environment, but also offers flexibility in integrating new or evolved requirements into decision-making. Moreover, the agent reasons over environment feedback to deduce potential requirements that may remain undefined or vague due to dynamic factors and incomplete domain knowledge. Also, the unsatisfiability feedback improves the convergence of agent's policy update through steering it to satisfiable regions. 

Autonomous defense agents for CPSs will highly need to adopt the CSRL approach in order to avoid safety and security violations. 
CPS usually exhibits many safety and security requirements that defense action must not violate to maintain the expected behavior of the infrastructure. We develop a use case scenario that simulates a CPS environment to assess the presented agent architecture. We show in our experiments that our agent converges to optimal planning at reasonable time windows despite having no prior knowledge, and the trained agent defeats attackers with diversified strategies within few time-sequences in 99\% cases. Hence, the outcome of our evaluation demonstrates the applicability of our agent for real-world cyber applications.     
\vspace{-1em}
\section{Overview of Integrated Reasoning Framework for Autonomous Agent}
\begin{figure}
    \centering
    \vspace{-1.6em}
    \includegraphics[height=4.3cm,width=10cm]{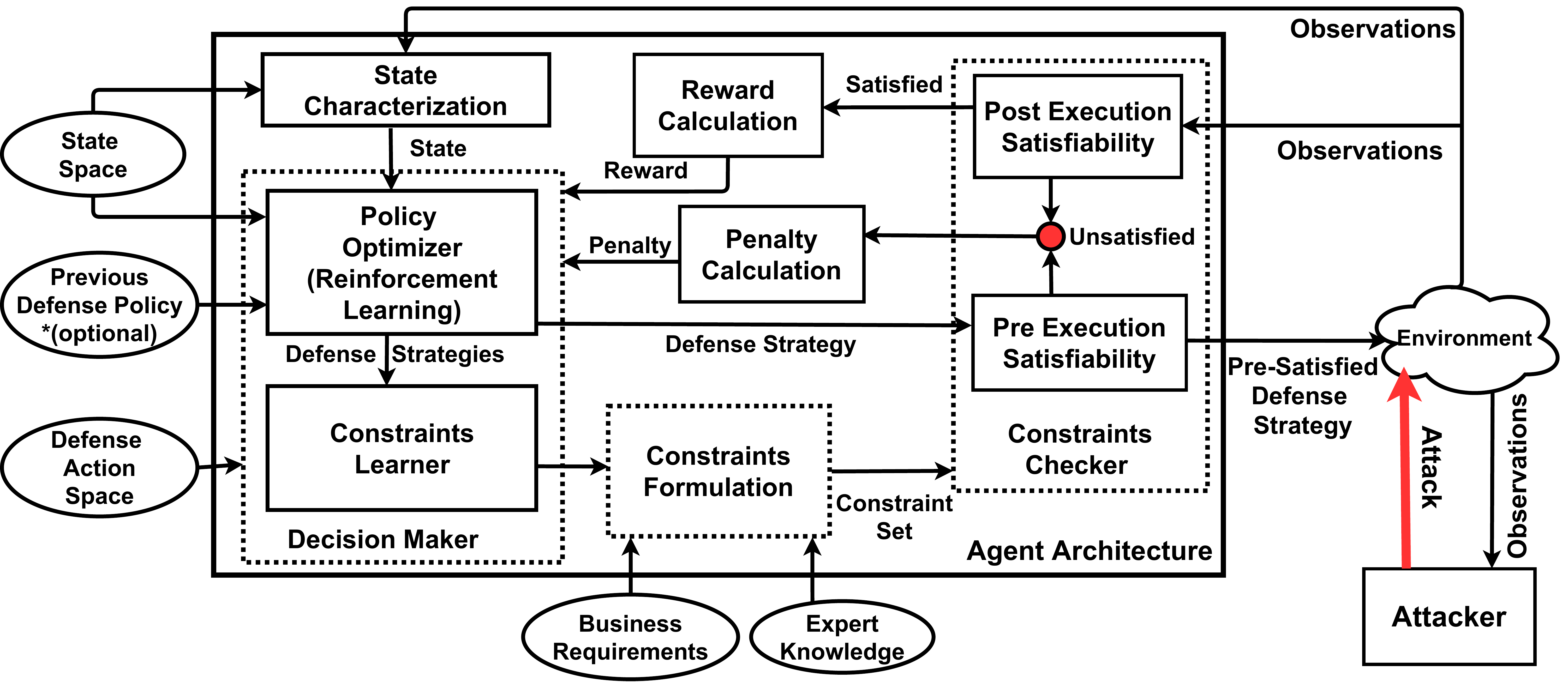}
	\caption{Autonomous Agent Architecture and Workflow. The environment contains an \textit{Attacker} who observes the environment and strategically executes attack actions. \textbf{Note\#} An arrow (input) ending at dotted box specifies that all modules inside that box receives the input.} 
	\label{fig:arch}
	 \vspace{-.25in}
\end{figure}

\vspace{-0.5em}
Fig. \ref{fig:arch} illustrates our framework that takes the State Space $S$ (set of possible states), Defense Action Space $A$ (set of possible defense actions), and optional previous policy (if any) as inputs. The \textit{Constraint Formulation} (cons-form) module composes initial Constraint Set by formulating known business requirements or expert knowledge. 
At the start of time-sequence $t$, \textit{State Characterization} module characterizes the current observation to a state (distinct environment condition) and sends to both \textit{Policy Optimizer} and \textit{Constraints Learner} (cons-learner). For the given state, \textit{Policy Optimizer} recommends the optimal action to \textit{Constraints Learner} and \textit{Pre Execution Satisfiablity} (pre-sat) modules. 

The pre-sat module checks if the recommended action can be deployed without violating any constraint of a specific subset of constraint set (i.e., received from cons-form module). If it fails, the agent sends a Penalty to the policy optimizer. The  optimizer updates the policy immediately based on the penalty and recommends new action during the same time $t$. Notably, the state remains unchanged due to not executing the action. In contrary, if the action satisfies all constraints, the agent executes it on environment as Pre-Satisfied Defense Action. In response or concurrently, the attacker executes his next attack action. 

Such attack and defense interplays trigger observations, based on which, the agent infers the action impact. Then, it checks whether the impact conforms with dynamic or remaining set of requirements (unchecked at pre-sat) at \textit{Post Execution Satisfiability} (post-sat) module. If it does not satisfy, the agent sends a Penalty to the optimizer; otherwise, it quantifies the rewards to send to the optimizer. The policy optimizer updates the policy based on these rewards or penalties. Moreover, based on such interactive experience involving action execution and receiving feedback, the agent's cons-learner learns or deduces new requirements that are sent to cons-form module to compose new constraint set.
\vspace{-1em}
\section{Autonomous Agent Architecture}
\vspace{-.5em}
This section describes components of Autonomous Agent architecture at Fig. \ref{fig:arch}.
\vspace{-2em}
\subsection{Constraints Formulation}
\label{subsec:cons_form}
Generally, cyber infrastructures such as CPSs contain diversified requirements that the computed defense strategy must not violate to maintain its expected behavior, safety, and security. These requirements can be business and mission-oriented, expert knowledge (e.g., historical experience), and others. Notably, expert knowledge may include commands of the network administrator, for example, keeping at least 20\% free resources to avoid hardware failures.  
The \textit{Constraint Formulation} (cons-form) module in Fig. \ref{fig:arch} formulates all such requirements as SMT constraints. 

Alongside user-given business requirements and expert knowledge, this module updates the constraint set when the agent learns new requirements based on its interactions with environment. Moreover, it modifies the set due to change of any business or existing requirements. Therefore, by this module, the agent's decision optimization can easily cope with the evolvement of requirements.


\vspace{-1em}
\subsection{Constraints Learner}
\label{subsec:cons_learner}
\vspace{-0.2em}
It is generally infeasible to know all constraints initially due to lack of deep domain knowledge or data, whereas some requirements can only be known after going into the operations due to environmental variability \cite{krismayer2019constraint}. 
For example, defining constraints for power system state estimation requires determining confidence on measured data. However, such domain-specific confidence depending on the likelihood of errors or sensor failures can only be computed by analyzing operational behaviors. 
Besides, deterministic approaches to specify such uncertain behaviors tend to be overly conservative.
Hence, critical infrastructures such as autonomous vehicles, smart grid nowadays endeavor to learn behavioral information from the environment. 

Our agent \textit{actively} learns new requirements using feedback (i.e., rewards, observations) of environment. 
In Fig. \ref{fig:arch}, the \textit{Constraints Learner} (cons-learner) module receives rewards or penalty as consequences of recently recommended defense actions. By analyzing rewards achieved at specific states, the agent deduces which group of actions should be avoided or preferred at particular environment conditions. For instance, if termination of a specific communication always induces intolerable business loss, the agent easily understands that the communication should remain untouched. However, before concluding observations to any such constraint, the agent must observe consequences of that action for multiple similar events due to non-deterministic environment behavior. Though we consider a static value for that required number of events, we plan to determine it dynamically in future extension. Moreover, there are ongoing research efforts to mine such constraints at real-time from observations such as runtime event logs or physical world information \cite{krismayer2019constraint}. 

\vspace{-1em}
\subsection{Constraints Checker}
\label{subsec:cons_checker}
\vspace{-0.2em}
The \textit{Constraint Checker} (cons-checker) module (dotted box at Fig. \ref{fig:arch}) uses SMT to check whether the recent defense strategy, recommended by Policy Optimizer, satisfies all formulated constraints or not. 
Our approach detaches cons-checker from the Policy Optimizer, because incorporating all requirements explicitly into optimization algorithm not only hardens the convergence of optimal policies but also may induce computational infeasibility. Therefore, rather than considering constraints directly, the optimizer considers rewards/penalty, computed based on the satisfiability of current constraint set by recent defense actions. This module performs constraints satisfiability verification in the following two phases:
    \vspace{-1.4em}
    \subsubsection{(1) Pre Execution Satisfiability Checker:} 
    This module verifies if there is any planning that can implement the recommended defense action without violating any constraint of \textit{Pre-satisfiable} constraint set (pre-cons).
    For example, if the recommended action wants traffic monitoring at several critical links, it checks whether any monitoring plan can achieve that within affordable energy. Importantly, pre-cons either do not rely on uncertain and dynamic environment factors or consider uncertain factors probabilistically. For example, Smart Grid considers various critical packet/message delay constraints \cite{wang2013cyber} by predicting packet delay, because it cannot be determined certainly due to unanticipated network effects such as changes in load balancing, or hardware failures. 
    
    In Fig. \ref{fig:arch}, the \textit{Pre Satisfiability} (pre-sat) module checks the conformity of the recommended defense action with current pre-cons. 
    Based on the satisfiability of these constraints, two following cases appear:
    
        \textbf{(a) Not Satisfied:} If the recommended action fails to satisfy any constraint of pre-sat, the agent immediately sends a \textit{Penalty} to the policy optimizer without executing the action. This is unlike traditional reinforcement learning approaches that update the policy only after executing the action.
        
        \textbf{(b) Satisfied:} If the recommended action satisfies all pre-sat constraints, it is executed as \textit{Pre-Satisfied} Defense Action on the environment.
        
    Our approach of not executing unsatisfiable actions makes the agent's RL exploration (exploration of action impacts) more effective by (1) avoiding execution of irrelevant (ineffective for current condition) actions that induce disastrous impact on real environment, and (2) offering flexibility for more explorations.
    \vspace{-1em}
    \subsubsection{(2) Post Execution Satisfiability Checker} 
    This module checks the satisfiability of a subset of constraints, termed as \textit{Post-satisfiable} constraint set (post-cons), after executing the pre-satisfied defense action on the environment. It is beneficial for any cyber system with following properties:
    
        \textbf{1. Constraints with dynamic or uncertain factors:} Certain verification of these constraints demands interactions with the environment, because scrutinizing impacts of actions on these dynamic factors require executing them. Importantly, even though such a constraint may be satisfied \textit{probabilistically} at pre-sat module, the agent checks its satisfiability as post-cons.
        
        \textbf{2. Numerous Constraints:} Verifying all constraints at runtime before executing an action may not be feasible for ensuring real-time defense optimization. Hence, the decision framework can only verify subset of constraints to ensure bounded computational overhead, and the remaining constraints need to be verified after the action execution.

    After executing the action, the \textit{Post Satisfiability} (post-sat) module at Fig. \ref{fig:arch} receives observations from the environment, and checks if the impact of action conforms all post-cons.
    Based on satisfiability, following cases appear:
    
   \textbf{(a) Not Satisfied:} If the executed defense action cannot satisfy any of post-cons, the agent sends a \textit{Penalty} to the policy optimizer for that action.
   
   \textbf{(b) Satisfied:} If it satisfies all post-cons, the agent forwards the recent observations to \textit{Reward Calculation} for quantifying the action payoffs and impact. 

\vspace{-1.2em}
\subsection{Policy Optimizer}
\label{subsec:policy_opt}
\vspace{-.3em}
Policy optimizer optimizes defense policy by maximizing action payoffs (rewards), that recommends an optimal defense action for a particular state. Due to no or limited knowledge about the environment initially, the agent applies Reinforcement Learning (RL) that updates the defense policy based on rewards or penalty received as feedback  \cite{sutton2018reinforcement}. Besides exploiting the previous experience or knowledge, RL algorithms optimally explore the consequences of other unexplored actions (i.e., RL-exploration). Thus, our agent applying RL computes optimal policy through learning the environment based on interactive experience.

The agent defines the environment and interactions using \textit{State} space $S$, \textit{Observation} space $O$, \textit{Action} space $A$, and \textit{Reward} function $R$. As shown in Fig. \ref{fig:arch}, the Policy Optimizer recommends the defense action for the current state and receives feedback.
This module uses Proximal Policy Optimization (PPO) \cite{schulman2017proximal} as RL algorithm, that shows better performance for continuous control tasks with two advantages: (1) constraining  policy  update  within a small range to avoid drastic deviation from old policy, and (2) performing multiple epochs on same minibatch data \cite{schulman2017proximal}. The first advantage helps the agent to cope with sensor noises or errors, whereas the second one aids to cope with the delayed feedback. PPO optimizes a clipped surrogate objective function, $L^{CLIP}(\theta)$, using Eqn. \ref{eqn:surrogate_objective}.
\begin{equation}
    \label{eqn:surrogate_objective}
    L^{CLIP}(\theta) = \mathbb{E}_t[min(r_t(\theta)A_t,clip(r_t(\theta),1-\epsilon,1+\epsilon)A_t)]
    \vspace{-.6em}
\end{equation}

where, $\theta$ represents policy parameter, $\epsilon$ is clip-range hyper-parameter, and $\pi_{\theta}$ and $\pi_{old}$ represents new and old stochastic policies respectively. Moreover, $r_t(\theta)=\frac{\pi_{\theta}(a_t|s_t)}{\pi_{old}(a_t|s_t)}$ specifies the likelihood ratio, where $\pi_{\theta}(a_t|s_t)$ specifies the probability of executing $a_t$ at state $s_t$ by $\pi_{\theta}$. Notably, PPO clips $r_t(\theta)$ if outside of $[1-\epsilon,1+\epsilon]$ to restrain large update. It formulates the advantage function $A_t$ by Eqn. \ref{eqn:adv_func}, considering $V(s_{t+l})$ (i.e., expected reward of state $s_{t+l}$ at time $t+l$) as baseline value to lower variance.
\vspace{-0.7em}
\begin{equation}
    \label{eqn:adv_func}
    A_t=\sum_{l=0}^T\gamma^l(r_{t+l}+\gamma V(s_{t+l+1})-V(s_{t+l}))
    \vspace{-0.8em}
\end{equation}

where, $\gamma\in [0,1)$ is discount factor that weighs future value, $r_{t+1}$ is the current reward or penalty, and $T$ is decision-horizon length until $\gamma^l>0$. 

PPO applies Advantage Actor Critic (A2C) approach \cite{schulman2017proximal} to optimize $L^{CLIP}(\theta)$, where \textit{Critic} estimates $V(s_{t})$ of $s_t$, and \textit{Actor} optimizes the policy based on $A_t$.


\vspace{-1em}
\subsection{State Characterization}
State represents a distinct condition of the environment based on critical environmental factors. Based on recent observations, the agent characterizes the current environment condition to a particular state for deciding the next optimal action. Symptoms observed from the environment or network may reveal the underlying state certainly or partially. 

Importantly, most of model-free RL algorithms implicitly address uncertainties associated with observations due to a partial observability, which is unlike the explicit \textit{Belief} calculation (probabilistic inference of current state) in model-based SDP. The applied PPO algorithm for our policy optimzation uses the characterized observation to decide the next action.  

\vspace{-1em}
\subsection{Rewards and Penalty Calculator}
\label{subsec:reward}
\vspace{-0.5em}
Reward quantifies the payoff of a defense action $a_d$ and provides as feedback to the policy optimizer.
Understandably, higher reward to a action for a state bias the optimizer to select that action due to its objective of maximizing rewards. Our agent assigns two types of rewards to $a_d$: (1) Penalty if $a_d$ fails to satisfy any pre or post constraint, and (2) Reward otherwise. 

The \textit{Reward Calculation} module at Fig. \ref{fig:arch} uses current observations to quantify rewards (can also be negative) based on the current status of the environment, improvement or degradation of CPS performance, user feedback on offered services, defense cost (includes deployment cost and negative impact), and others.
For a stochastic environment, reward function depends on multiple uncertain factors, and the administrator may change the weight of certain parameters or introduce new parameters based on his/her refined knowledge or new data. 
Whereas, the \textit{Penalty Calculation} quantifies the Penalty based on severity of constraints violation.
\vspace{-1em}
\section{Evaluation}
\label{sec:eval}
\vspace{-.5em}
This section describes the setup of experiments that are conducted to assess the agent's performance and discusses these experiments' outcome.

\vspace{-1em}
\subsection{Experiment Setup}
This section describes the use case and simulation parameters of our experiment.

    \textbf{Use Case Scenario:} We consider a CPS (e.g., smart grid) setting that accommodates anomaly-based detectors, to monitor critical connections among heterogeneous devices and provide probabilistic \textit{risk scores} based on anomalous behavior.
    These detectors consume varying energy based on required computation, and all detectors cannot be enabled at a time due to limited energy. 
    A device's risk score is the mean of all scores provided by enabled detectors at its multiple connections considering same accuracy of all detectors.
    There are two terminating states: (1) \textit{Attack-goal-state} when the attacker compromise at least 50\% of all devices, and (2) \textit{Attack-end-state} when the agent removes the attacker from all compromised devices. 
    
    \textbf{Attack Model:} The attacker aims to reach the attack goal state by propagating from compromised devices to connected (neighbor) devices. 
    We consider three types of attackers: (1) Naive attacker who randomly explores $N$ compromised nodes to propagate, (2) Stealthy attacker who strategically selects $\frac{N}{2}$ compromised nodes to explore while generating lower risk scores, and (3) Aggressive attacker who is stealthy and can explore $N$ machines.
    
    \textbf{Agent's Objective:} The agent may restart and reimage a device if its risk score is above than a threshold. However, such threshold needs to be dynamically selected to balance the trade-off between false positive (benign device identified as compromised) and false negative (compromised devices identified as benign) rate, considering current attack strategies and the number of enabled detectors. Therefore, the agent aims to dynamically compute the optimal threshold to reimage compromised devices and optimally enable detectors for efficient monitoring after satisfying all constraints at real-time.

    \textbf{RL Model Primitives:} The agent's defense space $A$ includes 3 qualitative levels for \textit{increasing} or \textit{decreasing} anomaly threshold $\delta_d$ (6 actions) followed by reimaging, 3 levels for \textit{increasing} or \textit{decreasing} enabled detector ratio $f$ of a device (6 actions), \textit{reimaging} of devices, and \textit{do nothing}.
    The state space $S$ consists of distinct compositions of 6 qualitative ratio levels of compromised devices (e.g., less than 50\%) with 3 levels (e.g., low number) of enabled-detector (18 states), and 2 terminating states. Importantly, a device, compromised or not, can be known certainly only after reimaging it; hence, the state characterization based on currently observed risk scores is uncertain.
    The agent's reward function $R$ is formulated using the following equation:
    \vspace{-0.3em}
    \begin{equation}
        \vspace{-0.3em}
        \label{eqn:reward}
        R(s,a) = -b_r\times C_r - d_r\times C_i + H_t\times I_w - H_g\times C_v
    \end{equation}
    where, $b_r$ is benign (non-compromised) devices reimaged, $d_r$ is the number of reimaged devices, boolean $H_t=1$ if the attack ends, boolean $H_g=1$ if the attack reaches goal state, $C_r$, $C_i$, and $C_v$ are costs, and $I_w$ is the incentive.
    
    \textbf{Constraints:} Pre-cons contains two vital requirements: (1) bounded expected energy consumption at a time by enabled detectors, and (2) enabling at least $l$ detectors for each device. To clarify, for a recommended action such as lowly increase $f$, the agent verifies if any detector-subset can satisfy all constraints. As Post-cons, it checks whether (1) real energy consumption and (2) loss due to reimaging benign devices are within tolerable limits.

    \textbf{\textbf{Implementation:}} We use Python3.6 to implement the framework and attack model that generates numerous attack scenarios to train and test the agent. We consider two topologies: Topo 1 with 100 devices, and Topo 2 with 200 devices. Our detectors' risk score distributions for compromised devices follow power law, whose tails stretch towards lower ends with increased attack stealthiness.   
    We use OpenAI Gym \cite{openAI} to simulate the environment, and use PPO2 and MlpPolicy libraries of Stable Baselines \cite{s_bl} to
    implement PPO. 

\vspace{-1.4em}
\subsection{Results}
\vspace{-0.5em}
We investigate (1) how efficiently the agent defends diversified attack strategies, (2) how fast the agent converges to optimal defense planning, and (3) how much benefits the constraints satisfying module offers.
    
\begin{wrapfigure}{r}{0.46\textwidth}
           \vspace{-2.1em}
          \centering    
          \includegraphics[height=3.1cm,width=0.45\textwidth]{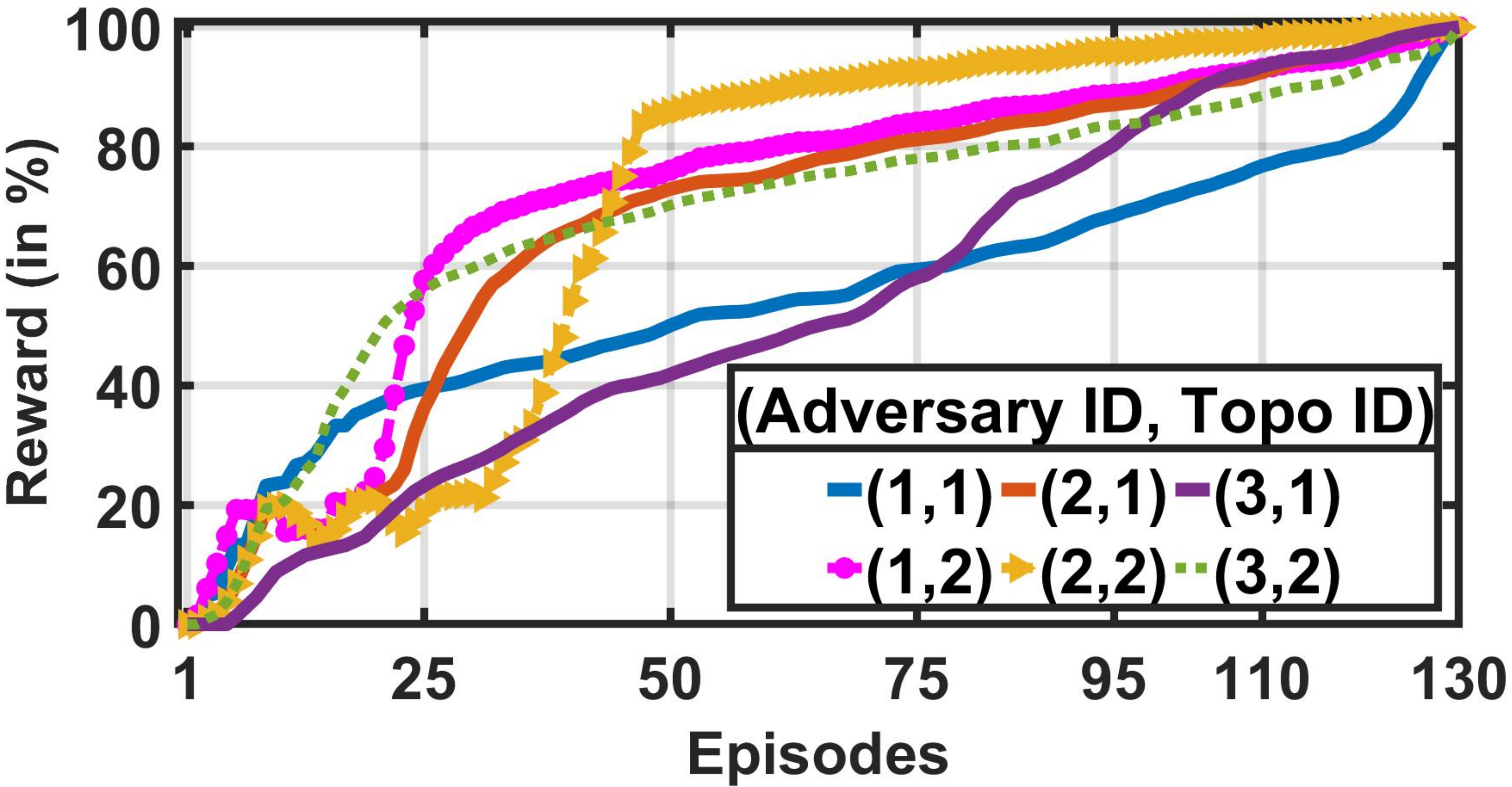}
          \vspace{-.1in}
          \caption{Reward (Normalized) w.r.t. training progression.}
          \label{fig:learning}
          \vspace{-.32in}
    \end{wrapfigure}

    \textbf{Agent's Learning Curve:} 
    Fig. \ref{fig:learning} illustrates the progression of agent's learning during training, where an Episode consists of 1000 time-sequences. 
    Here, for instance, rewards of plot (1,1) are normalized based on maximum reward achieved against Naive attacker at topo 1. As we can see, the agent converges faster for topo 2 despite slow start, due to more satisfiable plannings and opportunities to explore before termination state. Within 50 episodes, it reaches 87\% reward against Stealthy attacker (plot (2,2)), while plot (2,1) reaches only 68\% reward. Though convergences are slower against Aggressive attacker, the agent reaches more than 80\% rewards within 110 episodes in all scenarios.

            \begin{wrapfigure}{r}{0.46\textwidth}
            \vspace{-2.1em}
          \centering    
          \includegraphics[height=3.2cm,width=.45\textwidth]{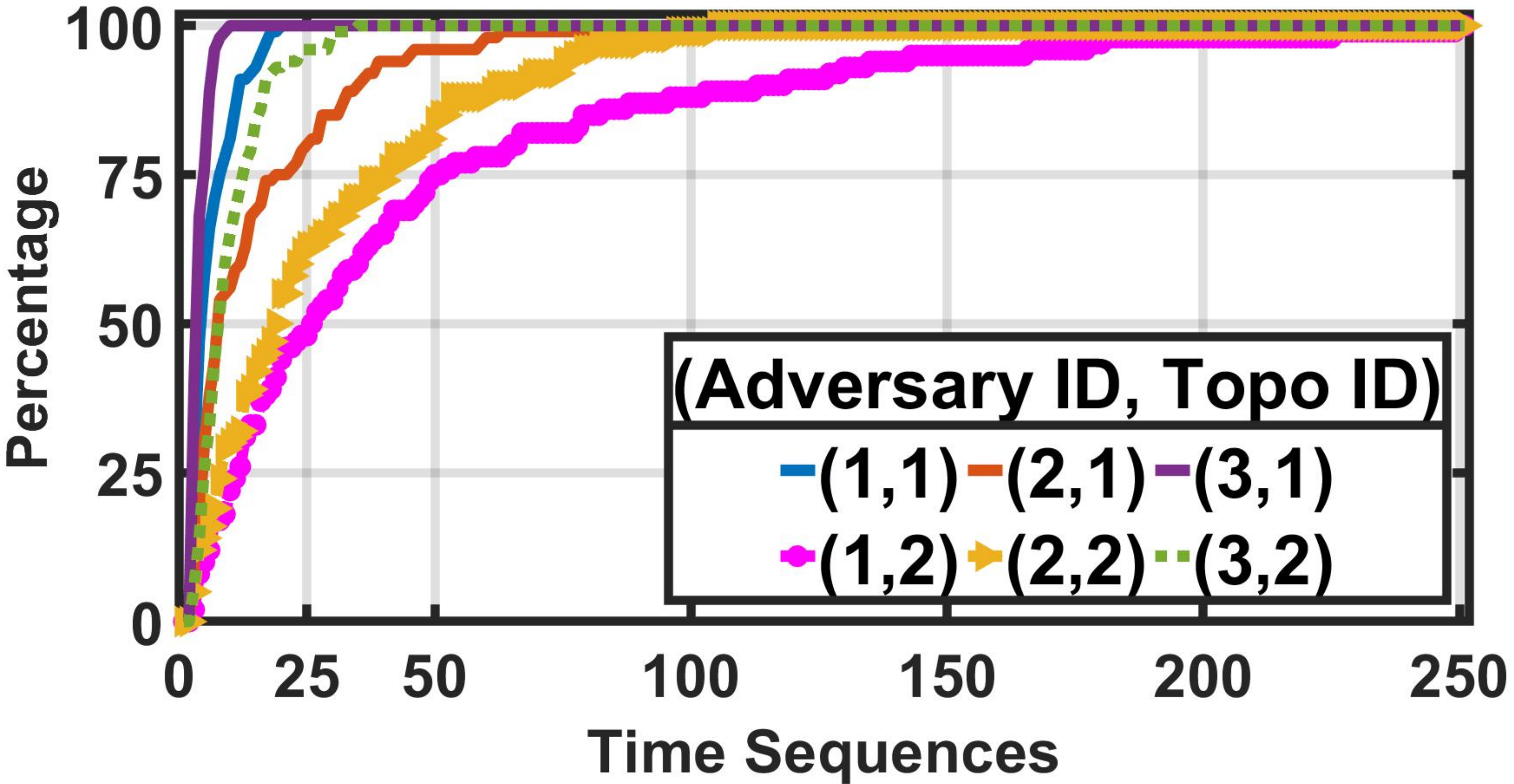}
          \vspace{-.1in}
          \caption{CDF of \textit{Required Time to reach Attack End State}.}
          \label{fig:cdf}
          \vspace{-0.3in}
    \end{wrapfigure}
    \textbf{Time to End Attack:} Fig. \ref{fig:cdf} shows a Cumulative Distribution Function (CDF) that describes how long the \textit{trained} agent takes to remove attacker from all devices during test settings. 
    For instance, a point (25,75) for plot (2,1) specifies that the agent stops the attacker propagation within 25 time sequences at 75\% cases. Importantly, the rate of attacker's reaching to \textit{Attack Goal State} is much lower than 1\%. 
    The agent terminates attack propagation within 100 time sequences in all cases except against Naive attacker at topo 2 whose distribution tail stretches until 175 time-sequences. It stops Aggressive attackers within 25-27 time sequences, while the stealthy attacker comparatively persists longer.
    
        \begin{wrapfigure}{r}{0.35\textwidth}
            \vspace{-3em}
          \centering    
          \includegraphics[height=2.4cm,width=.34\textwidth]{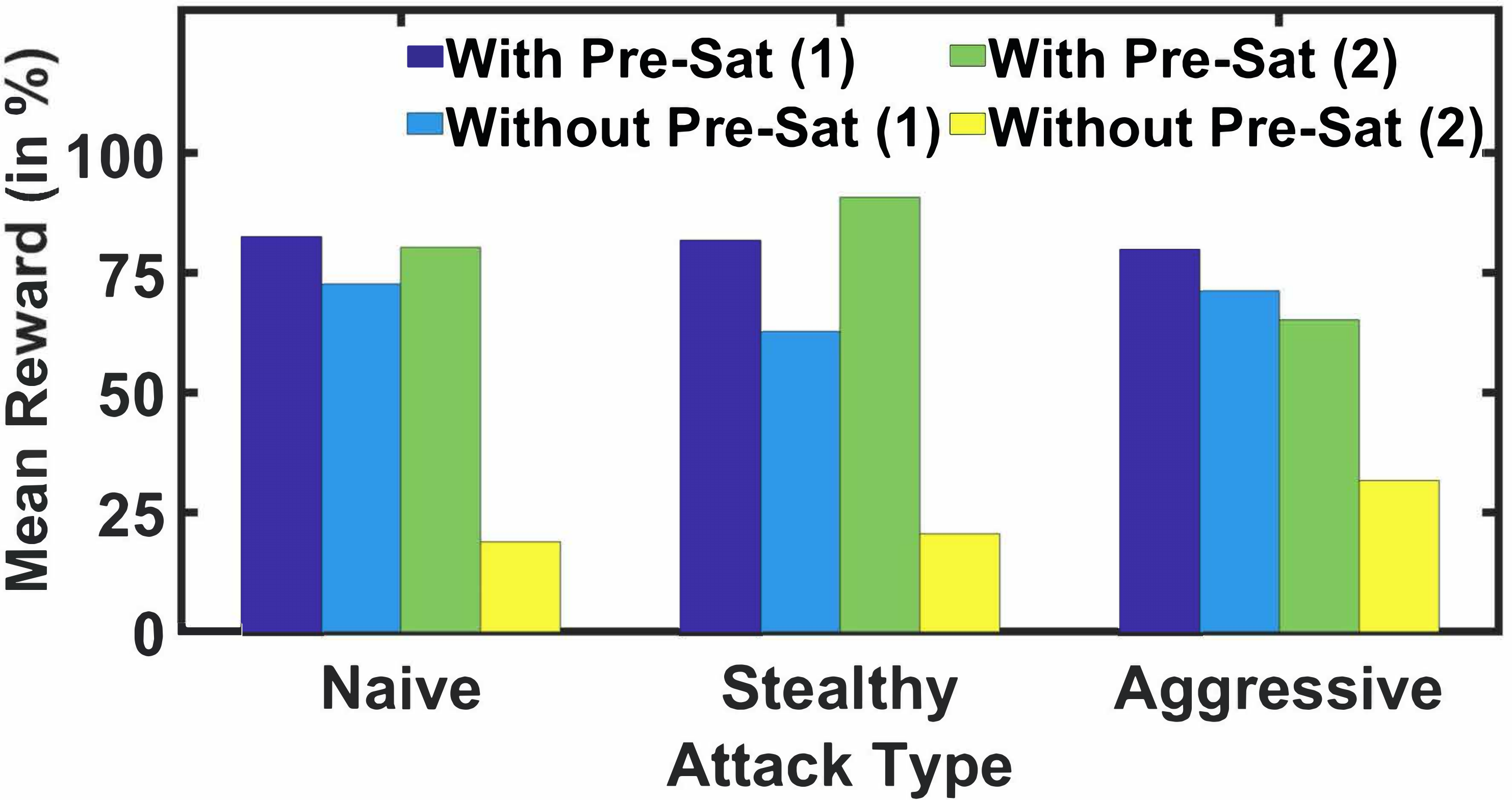}
          \vspace{-.1in}
          \caption{Mean Reward Comparison between approaches \textit{with} and \textit{without} Pre Execution Satisfiability.}
          \label{fig:cmp_cons}
          \vspace{-.5in}
    \end{wrapfigure}
    \textbf{Reward Comparison:} Fig. \ref{fig:cmp_cons} shows the benefit of Pre-sat module for topo 1 and 2, where rewards are normalized by the incentive  ($I_w$) of attack ending.
    The agent with Pre-sat always achieves more rewards, which is maximum (70\%) against Stealthy attacker and minimum (17\%) against Naive attacker at topo 2.
    Interestingly, though the agent terminates Aggressive attacker faster (at Fig. \ref{fig:cdf}), it executes comparatively expensive actions to defend them.

\vspace{-1em}
\section{Conclusion and Future Directions}
\vspace{-.5em}

Optimizing defense policies dynamically is a challenging tasks due to uncertainties of environment, strategical and adaptive attacks, and various safety and security requirements. In this paper, we present an architecture of Autonomous Defense Agent that optimizes defense planning at real-time using model-free Reinforcement Learning, while guaranteeing satisfaction of all requirements  using SMT-based constraints satisfiability verification. 
Moreover, our agent reasons over environmental observations to deduce new requirements and learn defense consequences. Our evaluation shows that our trained agent can defeat diversified attack strategies efficiently without requiring prior deep knowledge. Our approach is flexible to incorporate new and modified requirements easily into decision-making, and offers better scalability for real-time defense optimization in a constrained stochastic environment with dynamic or uncertain properties.

This architecture creates many interesting future research directions. First, our agent now learns new requirements based on rewards, but it will be interesting to find out how automated approaches can be developed to learn new requirements from network symptoms (e.g., logs, packets traces, and others). Besides, it is important to understand how much confidence the agent should at least have before introducing any new requirement. Second, defense payoffs may not always be observed immediately, and feedback such as user-complains may arrive several days later. We plan to investigate approaches to integrate the likelihood of such delayed feedback efficiently into policy optimization. Third, we would like to assess the scalability of the agent for higher dimensions of requirements, state space, and defense space of real-world applications. 

\vspace{-1em}
\bibliographystyle{plain}
\bibliography{rl}
\end{document}